\documentclass[a4paper, 12pt]{article}
\usepackage[T1]{fontenc}
\usepackage[english]{babel}
\begin{document}
\title{Evolution of interfaces and expansion in width
\thanks{Paper supported in part by KBN grant 2P03B 095 
13}     }
\author{H. Arod\'z and R. Pe\l{}ka \\
Institute of Physics, Jagellonian University, \\
Reymonta 4, 30-059 Cracow, Poland}

\maketitle
\begin{abstract}
Interfaces in a model with a single, real nonconserved order
parameter and purely dissipative evolution equation are considered.
We show that a systematic perturbative approach, called the expansion in 
width and developed for curved domain walls, can be generalized to 
the interfaces. Procedure for calculating curvature
corrections is described. We also derive formulas for local 
velocity and local surface tension of the interface. As an example,
evolution of spherical interfaces is discussed, including an estimate of 
critical size of small droplets.  
\end{abstract}
\noindent
PACS numbers: 61.30.Jf, 11.27.+d

\pagebreak

\section{Introduction}
Important aspect of dynamics of phase transitions in condensed matter is 
time evolution of an interface separating a retreating phase from the new one.
Studied in the framework of Ginzburg-Landau
type effective macroscopic models the interface 
can be regarded as a kind of smooth, asymmetric domain wall 
subject to a transverse force. The asymmetry and the force are due to a 
difference of potential energy across the interface.
Pertinent evolution equations 
for order parameters typically are nonlinear partial differential equations. 
In general they imply rather nontrivial phase ordering dynamics,
see, e.g., review article \cite{1}. Relativistic version of the problem, not
considered here, also is interesting because of its 
connection with field-theoretical cosmology, \cite{2}.     

Recently, evolution of ordinary domain walls has been studied with the help
of Hilbert-Chapman-Enskog method applied in
a suitably chosen comoving coordinate system \cite{3}, \cite{4}, \cite{5} -- 
a systematic and consistent perturbative scheme has been developed.
It yields the relevant solutions of the evolution equations in the form of 
expansion in a parameter $l_0$ which can be 
regarded as a measure of width of a static planar domain wall. Consecutive
terms in this expansion contain extrinsic curvatures of a surface 
comoving with the wall, and also 
certain functions (below denoted by $C_k$) which 
can be regarded as fields defined on that surface and coupled to the 
extrinsic curvatures. In the present paper, which is a sequel
to the previous paper \cite{3},  we show that that 
perturbative expansion can be generalised to the case of curved interfaces. 

The Hilbert-Chapman-Enskog method and the comoving coordinates
technique, which we have learned from \cite{6}, \cite{7}, respectively, 
have already been used in theoretical investigations of planar interfaces 
\cite{6}, and of curved ones in superconducting films \cite{8}. We apply 
these tools to curved interfaces in the three dimensional space, in a
Ginzburg-Landau type model defined by formulas (1), (2) and Eq.(3) below. 
Interfaces in this model have been investigated for a
long time, see, e.g., \cite{9}. 
Our main contribution consists in providing a systematic iterative scheme for 
generating the relevant solution in the form of a perturbative series.
The role of small parameters is played by the ratios $l_0/R_i,$ where  $l_0$ 
is the width and $R_i$ curvature radia of the interface. In spite of
nonlinearity of the evolution equation the perturbative contributions can 
be generated in surprisingly simple
manner. This is achieved by introducing the functions $C_k$ which
saturate certain integrability conditions. As an application, we derive 
a formula for local velocity of the interface with curvature
corrections included, and we discuss critical size of nucleating 
spherical droplets.

We consider a system described by a real, scalar, non-conserved order 
parameter $\Phi$, with the free energy $F$ of the form
\begin{equation}
F =  \int d^3x \left( \frac{1}{2}
K \frac{\partial \Phi}{\partial x^{\alpha}}
\frac{\partial \Phi}{\partial x^{\alpha}}  + V(\Phi) \right),
\end{equation}
where
\begin{equation}
V = A \Phi^2 + B \Phi^3 +C\Phi^4.
\end{equation}
Time evolution is governed by the dissipative nonlinear equation
\begin{equation}
\gamma \frac{\partial \Phi(\vec{x},t)}{\partial t} = 
 K \Delta \Phi - V'(\Phi).
\end{equation}
Here $(x^{\alpha})_{\alpha=1,2,3}$ are Cartesian coordinates in the space,
$V'$ denotes the derivative $dV/d\Phi$, and $K, \gamma, A, B, C$ are 
positive constants. The free energy of the form (1) arises in, e.g.,
de Gennes-Landau description of nematic-isotropic transition in nematic 
liquid crystals in a single elastic constant approximation ($L_2=0$) 
\cite{10}. Then $\Phi=0$ corresponds to the isotropic liquid phase, while 
in the nematic phase $\Phi\neq 0$.

The concrete form (2) of the potential has the
advantage that solution of Eq.(3) describing a planar interface has a simple,
explicitly known form. It can be found in, e.g., \cite{9}. The planar 
interface plays important role in the perturbative scheme: the main 
idea is that there exist curved
interfaces which do not differ much from the planar one if considered in 
appropriately chosen coordinate system (which in particular should comove 
with the interface). Therefore, one may hope that $\Phi(\vec{x},t)$ for such
curved interfaces can be calculated perturbatively, with the planar interface
giving the zeroth order term. To this end, it is necessary to introduce the
comoving coordinate system explicitly, and to give a prescription for the 
iterative computation of the perturbative corrections. 

The plan of our paper is as follows. In Section 2 we recall the planar
interface and the comoving coordinates. This Section contains preliminary
material quoted here for convenience of the reader as well as in order to 
fix our notation. In Section 3 we describe the perturbative scheme for
the curved interfaces. In Section 4 we present formulas for the 
velocity and the free energy of the curved interface. Section 5 is devoted to
a discussion of spherical droplets of the stable phase which nucleate during
the phase transition. Several remarks are collected in Section 6. 
In  Appendix A we construct solutions of linear equations obeyed by
corrections to transverse profile of the interface. Appendix B contains
a brief discussion of stability of the interface. 

\section{The preliminaries}
\subsection{The homogeneous planar interface}
Let us assume that the planar interface is perpendicular to the $z$ 
axis ($z\equiv x^3$) and homogeneous. Then $\Phi$ depends only on $z$ and $t$,
and Eq.(3) is reduced to 
\begin{equation}
\gamma \partial_t\Phi = K \partial^2_z\Phi - V'(\Phi). 
\end{equation}
The interface type solution 
$\Phi(z,t)$ interpolates smoothly between minima of $V$ when $z$ changes from 
$-\infty$ to $+\infty$. $V$ given by formula (2) has two minima 
\[ \Phi_-=0, \;\; \Phi_+= -\sqrt{K/(8Cl_0^2)},  \] 
where $l_0$ is given by formula (9) below.  The corresponding phases
we shall call isotropic and ordered, respectively.
Let us multiply Eq. (4) by $\partial_z\Phi$ and integrate over $z$.
The resulting identity 
\[ \gamma \int^{+\infty}_{-\infty} dz \partial_z\Phi \partial_t\Phi =
V(\Phi_-) - V(\Phi_+),  \]
implies that  $\partial_t\Phi\neq0$ if the minima are nondegenerate.
Further assumption that the interface moves in a uniform manner with velocity
$v_0$, that is that
\[ \Phi(z, t) = \Phi(z-z_0-v_0t),  \]
leads to the formula
\begin{equation}
\gamma v_0 \int^{+\infty}_{-\infty} dZ \Phi^2(Z) = V(\Phi_+) - V(\Phi_-),
\end{equation}
where
\begin{equation}
Z= z - z_0 - v_0 t. 
\end{equation}
Hence the interface moves towards the region of higher potential 
$V(\Phi_{\pm})$, as expected. 

The exact solution of Eq.(4) has the following form \cite{9}
\begin{equation}
\Phi_0 = - \sqrt{\frac{K}{8l_0^2C}} \frac{1}{1+ \exp(- Z/2l_0)},
\end{equation}
where
\begin{equation}
v_0=\sqrt{\frac{9K}{32 \gamma^2 C}} \left( B - \sqrt{9B^2 - 32 AC}\right),
\end{equation}
and
\begin{equation}
l_0^{-1} = \frac{1}{2\sqrt{2KC}}\left(3B + \sqrt{9B^2 - 32 AC}\right).
\end{equation}
The  constant $z_0$ can be 
regarded as the position of the interface at $t=0,$ and $l_0$ as its
width. $\Phi_0$ smoothly interpolates between the local minima of $V$:
$\Phi_-$ for $z \rightarrow -\infty$ 
and $\Phi_+ $ for $z \rightarrow
+\infty$. The corresponding values of the potential are
\[ V(\Phi_-)=0, \;\; V(\Phi_+) = \frac{K\gamma v_0}{96 l_0^3 C}. \]
At $\Phi_m=-3B/4C -\Phi_+$ the potential V has a local maximum if $A>0$.
The substitutions $Z \rightarrow -Z$ and $v_0 \rightarrow - v_0$ in 
formulas (7), (8) give another solution of Eq.(4), called 
anti-interface.

It is clear that solution (7) exists if 
\begin{equation}
9B^2 \geq 32 AC.
\end{equation}
The parameter $A$ has the following dependence on the temperature $T$ 
\cite{10}
\[ A = a (T- T_*), \]
where $a >0.$ The constants $B,C$ and $a$, approximately do not depend on 
temperature. Condition (10) is satisfied if the
temperature $T$ is from the interval $ (T_*, T_c)$, where $T_c$ is determined 
from the equation  $9B^2 = 32 a C (T_c-T_*)$. It is clear that $T_c >T_*$.
For temperatures in this interval one phase is stable and the other 
one is metastable.

The  potential (2) can also lead to a static, symmetric domain wall. 
Namely, for the 
temperature $T_0$ such that $B^2=4AC$ the velocity $v_0$ vanishes,
$V(\Phi_+) = V(\Phi_-) =0$, and the potential can be written in the 
following form
\begin{equation}
V=C \left[ (\Phi - \Phi_m)^2 - \Phi_m^2\right]^2,
\end{equation}
where for $T=T_0$
\[ \Phi_m = - \frac{B}{4C}.  \]
In this particular case there is the degenerate ground state given by 
$\Phi=\Phi_{\pm}.$ The potential (11) possesses the  $Z_2$ symmetry 
\[ \Phi \rightarrow 2\Phi_m - \Phi,  \]
and the interface becomes a static homogeneous, symmetric
domain wall with the $Z_2$ topological charge.

\subsection{The comoving coordinates}

Here we quote the main definitions
in order to introduce our notation. More detailed description of this 
change of coordinates, as well as a discussion of related mathematical
questions,  can be found in \cite{3}, \cite{4}, \cite{5}.

The comoving coordinates in the space, denoted by 
$(\sigma^{\alpha}) = (\sigma^1, \sigma^2, \sigma^3= \xi)$ where
$\alpha = 1, 2, 3$, are defined by the following formula
\begin{equation}
\vec{x} = \vec{X}(\sigma^i,t) + \xi \vec{p}(\sigma^i,t).
\end{equation}
Here $\vec{x}=(x^{\alpha})$, where $x^{\alpha}$ 
are the  Cartesian coordinates in the space $R^3$.
The points $\vec{X}(\sigma^i,t)$ form a smooth surface $S$
which is parametrised by the two coordinates  
\footnote{The Greek indices $\alpha, \beta,...$ have values 1,2,3 and they
refer to the threedimensional space, while the Latin indices
$i,j,k,l,...$ have values 1,2 and they refer to the inner coordinates
$\sigma^1, \sigma^2$ on the surface $S$.}
$\sigma^1, \sigma^2$.
In general, the interface moves in space, hence $\vec{X}$ depends on the  
time $t$. 

The surface $S$ is 
fastened to the interface --  the shape of it mimics the shape of the 
interface and they move together. We shall see that for
consistency of the perturbative scheme $\vec{X}(\sigma^i,t)$ has to obey 
certain equation from which one can 
determine evolution of the surface $S$. 
 The coordinate $\xi$ parametrises the axis  
perpendicular to the interface at the point $\vec{X}(\sigma^i,t)$. 
The vector $\vec{p}(\sigma^i,t)$  is a unit normal to $S$ at this point,
that is 
\[ \vec{p}^{\;2} =1, \;\; \vec{p}\; \vec{X}_{,k}
(\sigma^i,t) =0,   \]
where $ \vec{X}_{,k} =\partial\vec{X}/\partial \sigma^k$.
The surface $S$ is characterized in particular by  
induced metric tensor on $S$
\[ g_{ik} = \vec{X}_{,i} \vec{X}_{,k},   \]
and the extrinsic curvature coefficients 
\[ K_{ij} = \vec{p}\vec{X}_{,ij}. \]
The matrix $(g^{ik})$ is by definition the inverse of the 
matrix $(g_{kl})$, i.e. $g^{ik}g_{kl}= \delta^i_l$.

The two by two matrix $(K_{ik})$ is
symmetric. Two eigenvalues  $k_1,k_2$ of the matrix $(K^i_j)$, where 
$K^i_j = g^{il}K_{lj}$, are called extrinsic curvatures of $S$ at the 
point $\vec{X}$. The main curvature radia are defined as $R_i =1/k_i$. 
Thus, by the definition
\[ K^i_i = \frac{1}{R_1} + \frac{1}{R_2}, \;\; det(K^i_j)=\frac{1}{R_1R_2}. \]
In general the curvature radia vary along $S$ and with time.

The coordinates 
$(\sigma^{\alpha})$  replace the Cartesian coordinates 
$(x^{\alpha})$ in a vicinity of the interface. Components of metric tensor  
in the space transformed to the new coordinates are denoted 
by $G_{\alpha \beta}$. 
They are given by the following formulas
\[ G_{33} =1, \;\; G_{3k} = G_{k3} =0, \;\; G_{ik} = N^l_i g_{lr}N^r_k, \]
where 
\[ N^l_i = \delta^l_i - \xi K^l_i. \]
Dependence of $G_{\alpha \beta}$ on the transverse coordinate 
$\xi$ is explicit, and  $\sigma^1, \sigma^2$ enter through the tensors
$g_{ik}, \; K^l_r$ which characterise geometry of the surface $S$.

Components $G^{\alpha \beta}$ of the inverse metric tensor have the form  
\[ G^{33} =1,\;\; G^{3k}=G^{k3}=0, \;\; G^{ik}= (N^{-1})^i_r g^{rl} 
(N^{-1})^k_l, \]
where 
\[ (N^{-1})^i_r = \frac{1}{N} 
\left[(1-\xi K^l_l) \delta^i_r + \xi K^i_r \right], \]
and
\[ N = det(N^i_k) = 
1 - \xi K^i_i + \frac{1}{2} \xi^2 (K_i^i K^l_l - K^i_lK^l_i) 
= (1-\frac{\xi}{R_1})(1-\frac{\xi}{R_2}). \]

In order to transform Eq.(3) to the comoving coordinates we use the standard
formula 
\begin{equation}
\Delta\Phi = \frac{1}{\sqrt{G}} \frac{\partial}{\partial\sigma^{\alpha}}
\left(\sqrt{G} G^{\alpha\beta} \frac{\partial \Phi}{\partial \sigma^{\beta}}
\right),
\end{equation} 
where $G=det(G_{\alpha\beta}),\;  \sqrt{G} =\sqrt{g} N,\; g=det(g_{i k})$.

The time derivative in Eq.(3) is taken under the condition 
that all $x^{\alpha}$ are constant. It is convenient to use time derivative 
taken at constant $\sigma^{\alpha}$. They are related by 
the formula
\begin{equation}
\frac{\partial}{\partial t}|_{x^{\alpha}} = 
\frac{\partial}{\partial t}|_{\sigma^{\alpha}} +
\frac{\partial \sigma^{\beta}}{\partial t}|_{x^{\alpha}}
\frac{\partial}{\partial \sigma^{\beta}}.
\end{equation}

Finally, let us  introduce the dimensionless variables $s$
and $\phi$ instead of, respectively,  $\xi$ and $\Phi$:
\begin{equation}
\xi = 2l_0 s, \;\; \Phi(\xi, \sigma^i, t) = - \sqrt{\frac{K}{8Cl_0^2}}
\phi(s, \sigma^i,t). 
\end{equation} 
The coordinate $s$ gives the distance 
from $S$ in the 
unit $2l_0$ related to the width of the planar interface.

Using formulas (13-15) we can write  Eq.(3) in the following form, 
which is convenient for construction of the expansion in  width:
\begin{eqnarray}
& \frac{2 l_0^2\gamma}{K} 
\frac{\partial \phi}{\partial t}|_{\sigma^{k}} - 
\overline{v} \frac{\partial \phi}{\partial s} 
- \frac{2 l_0^2 \gamma}{K} (N^{-1})^i_k g^{kr} 
 \vec{X}_{,r} (\dot{\vec{X}} + 2 l_0 s
\dot{\vec{p}}) \phi_{,i} & \nonumber  \\
& = \frac{1}{2} \frac{\partial^2\phi}{\partial s^2} 
+ \frac{1}{2N} \frac{\partial N}{\partial s}
\frac{\partial \phi}{\partial s} + 2 l_0^2 \frac{1}{\sqrt{g} N} 
\frac{\partial}{\partial \sigma^j} \left(G^{jk}\sqrt{g} N 
 \phi_{,k} \right) &\\
& - \alpha \phi + (1+\alpha)\phi^2 - \phi^3,  & \nonumber 
\end{eqnarray}
where 
\[ \overline{v} = \frac{\gamma l_0}{K} \vec{p}\dot{\vec{X}} \]
is dimensionless transverse velocity of the surface $S$, 
the dot denotes the derivative $\partial /
\partial t|_{\sigma^{\alpha}}$, and
\[\alpha=\frac{4Al_0^2}{K}. \]
Formula (9) and the condition (10) imply that  $0 \leq \alpha \leq 1$ for 
temperatures in the range $[T_*, T_c]$.

The homogeneous planar interface (7) can be 
obtained from the evolution equation 
written in the form (16) in the following manner. 
As the surface $S$ we take a plane, hence $K^i_j =0$. Moreover, $S$ is
assumed to move with constant velocity $v_0$, hence
\[ \vec{p}\dot{\vec{X}}_0 = v_0 = \mbox{const}. \]
Finally,  
\[ \frac{\partial \phi}{\partial t}|_{\sigma^{\alpha}} =0      \]
because we look at the interface from the comoving reference frame, 
and $\partial \phi /\partial \sigma^i = 0$ because of the homogeneity.
Then equation (16) is reduced to
\begin{equation}
- \overline{v}_0  \frac{\partial\phi}{\partial s} = 
\frac{1}{2}\frac{\partial^2\phi}{\partial s^2} - \alpha \phi +(1+\alpha)\phi^2
-\phi^3.
\end{equation}
The solution previously given by formulas (6-9) now has the form
\begin{equation}
\phi = \phi_0(s), \;\;  \overline{v}_0 = \alpha - \frac{1}{2},
\end{equation}
where
\begin{equation}
\phi_0(s)=\frac{\exp(s-s_0)}{1+\exp(s-s_0)}.
\end{equation}
$\phi_0(s)$ smoothly interpolates between 0 and 1. This 
corresponds to interpolation between the minima $\Phi_-, \Phi_+$ of the 
potential $V$ if $\vec{p}$ is directed from
negative towards positive $z$'s. 
If we choose the opposite direction for $\vec{p}$ we obtain the 
anti-interface.
The constant $s_0$ corresponds to $z_0$ from formula (6).

\section{The expansion in width for curved interfaces}
Let us begin from a brief description of ideas underlying the calculations 
presented below. The set of solutions of the nonlinear, partial differential
equation (3) is very large. We are interested 
here only in a rather special subset of it, 
consisting of solutions which represent evolution of a smooth interface.
Moreover, even within this subclass we concentrate on rather special 
interfaces, called by us the `basic' ones. Their defining feature is that 
one can find a comoving coordinate system in which the order parameter of the 
interface is essentially given by $\phi_0(s)$, formula (19), modified 
by small corrections which take into account the nonvanishing
curvature. 

By writing the evolution equation in the form (16) we have shown that $l_0$
can be regarded as a parameter analogous to a coupling constant -- it appears
in Eq.(16) only as a coefficient in several (but not all) terms. Therefore, 
one may hope that a systematic perturbative expansion in $l_0$ will turn out 
useful, as it is the case with other perturbative expansions so numerous in 
theoretical physics. The perturbative series can be constructed in the 
standard manner: the sought for solution $\phi$ and the velocity 
$\overline{v}$ are written in the form 
\begin{equation}
\phi =\phi_0 + l_0 \phi_1 + l_0^2\phi_2 +... , \;\;  \overline{v} = 
\overline{v}_0 +l_0 \overline{v}_1 +l_0^2 \overline{v}_2 +... ,
\end{equation}
and inserted in Eq.(16). Coefficients in front of successive powers of $l_0$
in this equation are equated to zero. Notice that 
after the rescaling $\xi=2l_0s$ the expansion parameter $l_0$ is present 
also in $N$ and $(N^{-1})^i_k$. 
In the zeroth order we obtain the following equation
\begin{equation}
- \overline{v}_0  \frac{\partial\phi_0}{\partial s} = 
\frac{1}{2}\frac{\partial^2\phi_0}{\partial s^2} 
- \alpha \phi_0 + (1+\alpha)\phi_0^2 -\phi_0^3.
\end{equation}
which formally coincides with Eq.(17). Therefore, we can immediately write
the relevant solution
\begin{equation}
\overline{v}_0 = \alpha - \frac{1}{2}, \;\;
\phi_0(s, \sigma^i, t)=
\frac{\exp(s-C_0(\sigma^i,t))}{1+\exp(s-C_0(\sigma^i,t))}.
\end{equation}
There are however two differences between the planar solution (18), (19) 
and the solution (22). First, we do not assume homogeneity of the interface, 
therefore the constant $s_0$ from formula (19) is replaced by the function
$C_0(\sigma^i, t)$ of the indicated variables. Second, the surface $S$ is not
fixed yet, while in the former case it was a plane. 

It is convenient to rewrite Eq.(16) as equation for the corrections 
$\delta\phi, \delta\overline{v}$, which are defined by the formulas
\begin{equation}
\phi = \phi_0(s, \sigma^i, t) + \delta \phi, \;\; 
\overline{v} = \overline{v}_0 + \delta\overline{v}.
\end{equation}
After taking onto account the fact that $\phi_0$ obeys Eq.(21) we obtain 
equation of the form
\begin{equation}
\hat{L}\delta\phi = f,
\end{equation}
with the linear operator $\hat{L}$ 
\[
\hat{L} = \frac{1}{2}\frac{\partial^2}{\partial s^2} + (\alpha-\frac{1}{2})
\frac{\partial}{\partial s} - \alpha + 2(\alpha+1)\phi_0 - 3 \phi_0^2,
\]
and
\begin{eqnarray}
& f = - \left(\frac{1}{2N} \frac{\partial N}{\partial s} + \delta \overline{v}
\right) \frac{\partial\phi_0}{\partial s} +
\frac{2\gamma l_0^2}{K}
\left(\frac{\partial \delta \phi}{\partial t}|_{\sigma^{\alpha}}
- \frac{\partial C_0}{\partial t}|_{\sigma^{\alpha}}
 \frac{\partial\phi_0}{\partial s}\right)
 & \nonumber \\
&
- \frac{2 l_0^2 \gamma}{K} (N^{-1})^i_k g^{kr}
\vec{X}_{,r} (\dot{\vec{X}} 
+ 2 l_0 s \dot{\vec{p}}) \left(\delta\phi_{,i} -  C_{0,i} 
\frac{\partial\phi_0}{\partial s}\right)
&   \nonumber \\
& - \left(\frac{1}{2N} \frac{\partial N}{\partial s} 
+ \delta\overline{v}\right)
\frac{\partial\delta\phi}{\partial s} 
- 2 l_0^2 \frac{1}{\sqrt{g} N} 
\frac{\partial}{\partial \sigma^j} \left(G^{jk}\sqrt{g} N 
\left(\delta\phi_{,k} -  C_{0,k} 
\frac{\partial\phi_0}{\partial s}\right)\right)
& \nonumber \\
&+ (3\phi_0 -\alpha-1)(\delta\phi)^2 + (\delta\phi)^3,  &  
\end{eqnarray}
where $\overline{v}_0$ and $\phi_0$ are given by formulas (22).
Now it is easy to see that order by order 
in $l_0$ we obtain enumerated by $n=1,2,...$ inhomogeneous linear 
differential equations of the form 
\begin{equation}
\hat{L}\phi_n = f_n.
\end{equation}
For example, 
\begin{equation}
f_1= \left(K^i_i  - \overline{v}_1\right)
\frac{\partial \phi_0}{\partial s}.
\end{equation}
In the whole perturbative scheme Eq.(21) is the only nonlinear 
equation  for the contributions to the order parameter $\phi$. We show in 
the Appendix that Eqs.(26) can easily be solved with the help 
of standard methods -- one can construct the relevant Green's function
for $\hat{L}$. It is remarkable that the same operator $\hat{L}$ appears
in all equations (26), and that the form of it does not depend on the surface 
$S$. For these reasons calculation of the corrections $\phi_n$ is reduced
to relatively simple task of finding $f_n$ and calculating the one dimensional
integrals over $s$ shown in the Appendix. 

The perturbative Ansatz (20) and Eqs.(26) are two parts of the
Hilbert-Chapman-Enskog method. The third and most crucial part consists of 
integrability
conditions for Eqs.(26) \cite{6}. Such conditions appear because the operator
$\hat{L}^{\dagger}$, the Hermitean conjugate of $\hat{L}$, has a
normalizable eigenstate with the eigenvalue equal to zero. Such
eigenstate is called the zero mode. Let us first find the zero mode
for the operator $\hat{L}$: inserting $\phi_0$ in Eq.(21) and
differentiating this equation with respect to $s$ 
gives the following identity
\begin{equation}
\hat{L}\psi_r = 0,
\end{equation}
where 
\begin{equation}
\psi_r(s, \sigma^i, t) =
\frac{\partial \phi_0}{\partial s}
 = \frac{\exp(s- C_0)}{[1+\exp(s-C_0)]^2}.
\end{equation}
Notice that $\psi_r$ exponentially vanishes for 
$s\rightarrow \pm \infty$.
Because the operator $\partial/\partial s$ is
anti-Hermitean with respect to the scalar 
product $<g_1|g_2> = \int^{+\infty}_{-\infty} ds g_1^* g_2, $   the
operator $\hat{L}$ is not Hermitean. Its Hermitean conjugate has the form
\[ \hat{L}^{\dagger} = \frac{1}{2} \frac{\partial^2}{\partial s^2} -
(\alpha - \frac{1}{2}) \frac{\partial}{\partial s} - \alpha  + 2(\alpha +1)
\phi_0 - 3 \phi_0^2.     \]
The operator  $\hat{L}^{\dagger}$ has a zero mode too, namely
\begin{equation}
\hat{L}^{\dagger} \psi_l =0,  
\end{equation}
where
\footnote{The subscripts l and r stand for left and right, respectively. The 
point is that (30) can be written as $\psi_l\hat{L}=0$.}
\begin{equation}
\psi_l = \exp[(2\alpha -1)(s-C_0)]\psi_r.
\end{equation}
The function $\psi_l$ vanishes exponentially for 
$s \rightarrow \pm \infty$ because
$0 < \alpha <1 $ for all temperatures in the range  $(T_*, T_c)$.
For $\alpha=1/2$, that is when the interface becomes the 
domain wall, the two zero modes coincide. 

Let us multiply both sides of Eqs.(26) by $\psi_l(s)$ and take the integral
$\int^{+\infty}_{-\infty} ds$. The l.h.s. of the resulting formula vanishes
because of (30), hence
\begin{equation}
\int^{+\infty}_{-\infty} ds \; \psi_lf_n =0
\end{equation}
for $n=1,2,...$  .   It turns out that these conditions are nontrivial. In 
particular, they give evolution equation for the surface $S$. 

It should be noted that the
conditions (32) are in fact approximate, but the neglected terms are
exponentially small. The point is that in order to obtain Eqs.(26) we use
the expansions of the type
\[ \frac{1}{1-2l_0s/R_i} = \sum^{\infty}_{k=0}
\left(\frac{2l_0s}{R_i}\right)^k                         \]
$(i=1,2)$, which are convergent for $s<s_M$ where $s_M=\mbox{min}(R_1/2l_0,
R_2/2l_0)$. Therefore, when deriving conditions (32), the integration range
should be restricted to $|s|<s_M$. Because of the exponential decrease of
$\psi_l$ and $f_n$ at large $|s|$, this will give exponentially small
corrections to these conditions. We assume that the ratios $R_i/l_0$
are so large that we may neglect those corrections.

Let us compute the basic interface up to the order $l_0$. For $n=1$,
condition (32) gives 
\begin{equation}
\overline{v}_1 = \frac{1}{R_1} + \frac{1}{R_2}
\end{equation}
because  
\[ a_0(\alpha)= 
\int^{+\infty}_{-\infty} ds \psi_l \psi_r = B(2\alpha+1, 3-2\alpha) 
\]
does not vanish for $\alpha$ in the interval (0, 1). Here $B$
denotes the Euler beta function. 
The function $a_0(\alpha)$ has the symmetric "U" shape in the interval 
$\alpha\in[0, 1]$, with the minimum equal to 1/6 at $\alpha=1/2$, and the 
upper ends reaching the value 1/3 for $\alpha=0$ and 1.
The condition (33) implies that the
surface $S$ obeys the following evolution equation
\begin{equation}
\frac{\gamma}{K} \dot{\vec{X}}\vec{p} =\frac{2 \alpha -1}{2l_0} 
+  K^i_i.
\end{equation}
It formally coincides with the well-known Allen-Cahn equation \cite{11}.

Now Eq.(26) with $n=1$ is reduced to 
\begin{equation}
\hat{L} \phi_1 =0.
\end{equation}
It has the following solution which vanishes at $s \rightarrow \pm \infty $ 
\begin{equation}
\phi_1 = C_1(\sigma^i, t) \psi_r,
\end{equation}
where  $C_1$ is a smooth function of the indicated variables.

For $n=2$,  we obtain from formula (25), after taking into account the results
(33) and (36),
\begin{eqnarray}
& f_2 = 
( 2s K^i_j K^j_i - \overline{v}_2) \partial_s\phi_0 
+ (3\phi_0 - \alpha -1) C_1^2 (\partial_s \phi_0)^2 &  \nonumber \\
& + \frac{2\gamma}{K} \left( -\partial_t C_0 + g^{ik} 
\vec{X}_{,k}\dot{\vec{X}} C_{0,i} \right) \partial_s\phi_0
+ 2\frac{1}{\sqrt{g}} \frac{\partial}{\partial \sigma^i}\left(g^{ik}\sqrt{g}
C_{0,k}\partial_s\phi_0\right).  &
\end{eqnarray}
Straightforward integration over $s$ as in (32) can be a little bit 
cumbersome. This calculation can be significantly simplified with the help 
of the following identity
\begin{equation}
2 \int ds \psi_l [3\phi_0 - (\alpha+1)] \partial_s\phi_0\phi_n = \int ds
\partial_s\psi_l f_n, 
\end{equation}
where in the case at hand $n=1$.  Identity (38) is obtained from Eq.(26) by
differentiating both sides of it with respect to $s$, multiplying by $\psi_l$
and integrating over $s$, just as in the derivation of the integrability
conditions (32). 
The integrability condition gives
\begin{eqnarray}
&  a_1(\alpha) K^i_j K^j_i - \overline{v}_2 + 
\frac{2\gamma}{K} \left( -\partial_t C_0 
+ g^{ik} \vec{X}_{,k}\dot{\vec{X}}  C_{0,i} \right)
& \nonumber \\ 
&+ 2\Delta_2C_0 + (2\alpha-1) g^{ik}  C_{0,i} C_{0,k} + 
2 C_0 K^i_j K^j_i =0, &
\end{eqnarray}
where 
\[ \Delta_2= \frac{1}{\sqrt g} \frac{\partial }{\partial\sigma^i}\left(
\sqrt{g}g^{ik} \frac{\partial}{\partial\sigma^k} \right) \]
is the Laplacian on the surface $S$ and
\[
a_1(\alpha) =a_0(\alpha)^{-1} \frac{d a_0(\alpha)}{d \alpha}.
\]
For $\alpha$ from the interval [0,1] the function $a_1(\alpha)$ is almost 
linear. In particular, $a(0) = -3, a(1/2)=0, a(1)=3.$ 

The integrability condition (39) leaves the freedom of choosing whether we 
keep
nonvanishing $C_0$ or $\overline{v}_2$. It is clear that we can not put 
to zero both of them unless $K^i_j K^j_i=0$ (then $S$ is a plane). The
choice $C_0=0$  gives 
\begin{equation}
\overline{v}_2 =  a_1(\alpha) K^i_j K^j_i.
\end{equation}
This implies a correction to the Allen-Cahn equation of the form
\[ 
\frac{\gamma l_0}{K} \delta_1(\dot{\vec{X}}\vec{p}) = l_0^2 \overline{v}_2, 
\]
where $\delta_1(.)$ denotes the first order  correction to $\dot{\vec{X}}
\vec{p}$. In consequence, there will be a correction to the solution
$\vec{X}$ of the Allen-Cahn equation, and corrections to $g_{ik}$ and
$K_{ik}$. These corrections have to be taken into account when calculating
$f_k$ with $k\geq 3.$ It is clear that this version of the perturbative 
scheme is rather cumbersome.

On the other hand, if we put $\overline{v}_2 = 0$,
then the evolution of the surface $S$ is still governed by the relatively
simple Allen-Cahn equation (34).
The integrability condition (39) is now 
saturated by the function $C_0 $ -- it has the form of evolution equation 
for $C_0$, namely 
\begin{eqnarray}
&\frac{\gamma}{K}\left[\partial_t C_0 - g^{ik} \partial_{\sigma^k}
\vec{X}\dot{\vec{X}}  C_{0,i}\right]  
-  \Delta_2C_0  &  \\
& - (\alpha-\frac{1}{2}) g^{ik} C_{0,i}
 C_{0,k} - C_0 K^i_j K^j_i = 
\frac{1}{2} a_1(\alpha) K^i_j K^j_i. & \nonumber
\end{eqnarray}

Let us also check the third integrability condition. Formulas (20), (25) give
\begin{eqnarray}
& f_3 = -\overline{v}_3  \psi_r + 2 K_i^i(3K^i_jK^j_i - (K_i^i)^2) s^2 \psi_r
+\frac{2 \gamma}{K} \partial_t(C_1\psi_r) & \nonumber \\
& + \frac{2 \gamma}{K} g^{ir} \vec{X}_r \dot{\vec{X}}(C_1\psi_r)_{,i}
+\frac{4 \gamma}{K} C_{0,i}(g^{ir} \vec{X}_r \dot{\vec{p}}  K^{ir}
\vec{X}_r \dot{\vec{X}}) s \psi_r & \nonumber \\
& +2sK^i_jK^j_i \partial_s \phi_1 - 2\Delta_2(C_1\psi_r) + 4s K^i_i 
\frac{1}{\sqrt{g}}(\sqrt{g}g^{jk}C_{0,k}\psi_r)_{,j}  & \nonumber \\
&- \frac{4s}{\sqrt{g}} (\sqrt{g}g^{jk}K^l_l C_{0,k}\psi_r)_{,j} 
+ \frac{8s}{\sqrt{g}}(\sqrt{g}K^{jk}C_{0,k}\psi_r)_{,j} & \nonumber \\
& + 2(3\phi_0-\alpha -1) \phi_1 \phi_2 + \phi_1^3,  &  
\end{eqnarray}
where we have put $\overline{v}_2 =0$. The  term with  $\phi_2$ contains a 
new function $C_2$ (see the Appendix), however this function will not appear
in the integrability condition because the integration over $s$ eliminates 
the term proportional to $C_2$. This can be seen from the identity (38): on
the r.h.s. of it we have $f_2$ in which $C_2$ is not present. We see already 
from formula (42) that we can choose whether to keep nonvanishing 
$\overline{v}_3$ or $C_1$. For the same reason as in the case of $n=2$ we
choose $\overline{v}_3=0$, and the integrability is saturated by $C_1$. 
The resulting evolution equation for $C_1$ has the following form 
\begin{eqnarray}
& \frac{\gamma}{K}\left[\partial_t C_1 - g^{ik} \partial_{\sigma^k}
\vec{X}\dot{\vec{X}} \partial_{\sigma^i} C_1\right]  
- \Delta_2C_1  -  K^i_j K^j_i C_1      & \nonumber \\
& - (\alpha -\frac{1}{2}) g^{jk} \partial_k C_0 \partial_j C_1         = 
- \left(\frac{1}{4}a_2 + a_1C_0 + C_0^2\right) K^i_i [3 K^k_jK^j_k 
- (K^l_l)^2] & \nonumber \\
& + 2 \left[g^{ij} \partial_jK^l_l - \frac{\gamma}{K}[g^{ir}(\partial_r
\dot{\vec{X}}\dot{\vec{p}}) + K^{ir}(\partial_r\vec{X}\dot{\vec{X}})]\right]
\partial_i C_0 (C_0 + \frac{1}{2}a_1)  & \\
& - \frac{4}{\sqrt{g}}\partial_j(K^{jk}\sqrt{g} \partial_kC_0)(C_0 +
\frac{1}{2}a_1) - 2K^{jk}\partial_kC_0\partial_jC_0
\left[1+(2\alpha-1)(C_0+\frac{1}{2}a_1)\right], &
\nonumber 
\end{eqnarray}
where
\[ a_2 = \frac{a_0''}{a_0}.  \] 

It is easy to proceed to the second and higher orders. Using formulas from 
the Appendix  one can write general solution $\phi_n$ of Eq.(26). 
It contains the 
function $C_n(\sigma^i,t)$ which obeys an equation analogous to (41) or (43).
This equation follows from the integrability condition (32) with $n$ 
replaced by $n+2$ if we put $\overline{v}_n=0$. 
Due to identity (38), in the derivation of 
that equation we do not need explicit form of $\phi_{n+1}$. 
In the present paper we will stop our considerations
at the first order.

In order to obtain a concrete basic interface solution
we have to specify initial data for equations (34), (41), (43). There is no
restriction on the initial data, except the obvious requirement that 
perturbative corrections of given order should be small in comparison with
the ones from preceding orders. In particular, $l_0C_1 \ll 1, l_0^2C_2 \ll 1$
and $l_0/R_i \ll 1$. If we consider the interface solution only up to the
first order correction (36) then a simplification appears: without any loss 
in generality we may adopt the homogeneous initial data
\begin{equation}
C_0(\sigma^i, t=t_0) = 0, \;\; C_1(\sigma^i, t=t_0) = 0. 
\end{equation}
This can be justified as follows. Local deformation of $S$ by shifting
a small piece of it along 
the direction $\vec{p}$ results in the corresponding shift of the 
coordinate $s$. Therefore for any given basic interface we can choose 
initial position of $S$ such that  $C_0(\sigma^i, t=t_0) = 0$
in formula (22) for  $\phi_0$. This can be done at one instant of time,
e.g., at the initial time. Values of $C_0$ and position of $S$ at later times 
are determined uniquely by Eqs.(34), (41), and in general $C_0$ does not 
vanish. Notice however that such a shift wiil influence terms of the order 
$l_0^2$ and higher in formula (25) for $f$ -- due to the explicit presence of
$s$ in $N, (N^{-1})^i_k$ $f$ is not invariant under the translations
$s \rightarrow s+C_0$. 

The r.h.s. of Eq.(41) vanishes for $\alpha=1/2$, 
that is in the domain wall case. In this case the initial condition (44)
implies that $C_0 =0$ for all times, and in consequence $C_0$ disappears
from the first order perturbative solution. 
 
As for $C_1$, the reason for the homogeneous initial condition is that the 
interface with the first order correction, that is
\begin{equation}
\phi = \phi_0(s-C_0) + l_0 C_1 \psi_r(s-C_0),
\end{equation}
can be regarded as $\phi_0(s-C_0+l_0C_1)$ 
to the first order in $l_0$, so again
we can cancel $C_1$ at the initial time $t_0$ by suitably correcting the 
initial position of the surface $S$. Let us stress again that this works 
only at the fixed time instant. 

The initial data (44) imply that at the initial time the order parameter
$\phi$ is equal to  $\phi_0(s)$. Hence, the only freedom in choosing initial 
data for $\phi$ we still have is position of the surface $S$. When the second
and higher order corrections are included one has to allow for more general
than (44) initial data for $C_0$ and $C_1$, nevertheless the initial form
of $\phi$ always is uniquely fixed by these data and the initial position
of the surface $S$.

To recapitulate, the perturbative solution to the first order has the form 
(45) where $\phi_0, \psi_r$ are given by formulas (22), (29), respectively.
Formula (45) gives dependence on $s$ explicitly. The functions $C_0, C_1$
are to be determined from Eqs.(41),(43) with the initial data (44).
One also has to solve Allen-Cahn equation (34). In certain cases these 
equations can be solved analytically, e. g., for spherical
interface discussed in next Section.  In general case one will be forced
to use numerical methods. In comparison with the initial evolution equation
(3) the gain is that the equations for $S, C_0, C_1$ involve only two spatial
variables $\sigma^1, \sigma^2$. Such reduction of the number of independent
variables is a great simplification for numerical calculations.

The perturbative solution obtained above can be used in calculations of
physical characteristics of interfaces. In the following  Section 
we obtain formulas for local velocity and surface tension of the interface.
In the Section 5 we discuss evolution of a spherical interface. 

\section{Local velocity and surface tension of the interface}
Let us apply the expansion in width in order to find
local transverse velocity and surface tension of the interface. We shall
use the first order solution (45). The velocity is obtained 
from the condition $\phi=\mbox{const}$. It does not necessarily coincide with 
$\vec{p}\dot{\vec{X}}$ given by Allen-Cahn equation (34). Because we neglect
terms of second and higher order in $l_0$, we may write $\phi$ in 
the form $\phi_0(s-C_0 +l_0C_1)$ from which we see that $\phi$ is constant 
on surfaces given in the comoving coordinates by the condition 
$s-C_0+l_0C_1=s_0$, where $s_0$ is a constant. It follows from formula (12)
that in the laboratory Cartesian coordinate frame these surfaces
are given by  $\vec{x}_0(\sigma^i,t)$, where 
\[ \vec{x}_0(\sigma^i,t) = \vec{X}(\sigma^i,t) + 2l_0(s_0+C_0 -l_0C_1)
\vec{p}(\sigma^i,t).  \]
The transverse velocity of the interface is equal to $\dot{\vec{x}}_0\vec{p}.$
In order to calculate it we take time derivative of $\vec{x}_0$, 
project it on $\vec{p}$, and
use equations (34), (41), (43) with the initial data (44). 
The result can be written in the form
\begin{equation}
\frac{\gamma}{K} \vec{p}\dot{\vec{x}}_0 = \frac{2\alpha-1}{2l_0}
+ \frac{1}{R_1} +\frac{1}{R_2} + l_0 a_1
\left(\frac{1}{R_1^2} + \frac{1}{R_2^2}\right) + l_0^2 a_2 \left(
\frac{1}{R_1^3} +\frac{1}{R_2^3}\right).
\end{equation} 
The unit normal vector $\vec{p}$ is directed from the isotropic phase 
($\Phi = \Phi_0$) to the ordered phase ($\Phi=\Phi_+$).  
In the next Section we shall use formula (46) in the case of spherical 
droplets.

The surface tension is another basic characteristics of the interface.
It can be determined from a formula for free energy of the interface,
which is defined as follows. The surface $S$ cuts the total volume 
of the sample into two
regions denoted below by I and II. Let us imagine that in region I there is 
the homogeneous
isotropic phase with constant free energy density equal to $V(\Phi_-)=0$, 
and in region II the homogeneous ordered phase for which   
$V(\Phi_+)=K\gamma v_0/(96l_0^3C)$. The normal vector 
$\vec{p}$ points to region II. The free energy of the interface is defined as 
the difference
\[ F_i = F -  V(\Phi_+) {\cal V}_{II},     \]
where ${\cal V}_{II}$ denotes volume of the region II, and $F$ is the total
free energy of the sample given by formula (1).  
We shall compute $F_i$  using the first order solution (45). Then, without 
any loss of generality we may put $C_0=C_1=0$ at the given time, as argued 
in the preceding Section, while the surface $S$ remains arbitrary.
Therefore, we need only $\phi_0(s)= \exp(s)/(1+\exp(s))$. 
Because the dependence on the 
coordinate $s=\xi/2l_0$ is explicit, we can integrate over $s$ in the formula
(1) for the free energy $F$. The volume element and the gradient free energy 
are taken in the form
\[ d^3x = \sqrt{G} d\xi d\sigma^1 d\sigma^2, \;\; 
\frac{\partial\phi}{\partial x^{\alpha}} 
\frac{\partial\phi}{\partial x^{\alpha}} = G^{\alpha \beta}
\frac{\partial\phi}{\partial \sigma^{\alpha}} 
\frac{\partial\phi}{\partial \sigma^{\beta}}.   \]
Neglecting terms quadratic in $l_0/R_i$ we obtain the following formula 
for the free energy of the interface 
\[ F_i = \int_S \kappa \; dA,  \]
where 
\begin{equation}
\kappa =  \frac{K^2}{96l_0^3C} \left[ 1  
 -  (\frac{\pi^2}{3}-2)(1- 2\alpha) (\frac{l_0}{R_1}+
\frac{l_0}{R_2}) \right] 
\end{equation}
can be regarded as the local surface tension of the interface at the point  
$\sigma^1, \sigma^2$.
$R_1,R_2$ are the main curvature radia of the surface $S$ at that
point, and $dA=\sqrt{g}d\sigma^1d\sigma^2$ is the 
surface element of $S$. Of course, this formula for $F_i$ can be trusted if 
$l_0/R_i \ll 1$.  

For a spherical droplet of the ordered phase embedded in the isotropic phase
$R_1, R_2$ are positive (the signs follow from formulas given in Section 2.2)
and of course equal to the radius of the sphere. 
If $\alpha <1/2$, formula (46) implies that the droplet grows (if its radius 
is large enough) because $\vec{p}\dot{\vec{x}}_0 <0$ and $\vec{p}$ is the
inward normal. In this case the curvature correction 
diminishes the surface tension, and $\kappa$ increases as the droplet grows.
In the reverse situation -- the isotropic phase inside and the ordered one 
outside  -- the curvature increases the surface tension and $\kappa$ 
decreases as the droplet grows. 

In the case of a growing droplet of radius $R$ of the isotropic phase in
the ordered medium $\kappa$ has the same dependence on the curvature. Here
$\vec{p}$ is the outward normal, $\alpha >1/2,\; \vec{p}\dot{\vec{x}}_0 >0$,
and $R_1=R_2=-R$. Nevertheless the values of surface tension in both cases
are different because $l_0$ present in formula (47) depends on $\alpha$,
namely $l_0 \sim (1+\alpha)$, see formula (57) below. 

Notice that the first order curvature correction to $\kappa$ vanishes in the
domain wall case ($\alpha=1/2$).

\section{Evolution of spherical interface}
Let us now apply the formalism developed in Sections 3 and 4 to evolution of 
spherical droplets. We  assume that $\alpha \neq 1/2$ in order to exclude 
the relatively simpler case of the domain wall.
The surface $S$ is parametrised by 
\[
\vec{X}_0 = \mp R(t) \vec{p}(\theta,\psi). \]
Here $\theta, \psi$ are the spherical angles, and $\vec{p}$ is the inward 
(outward) normal to the sphere when $\alpha <1/2$ ($\alpha>1/2$). 
Thus,  $\vec{p}\dot{\vec{X}}_0 =\mp \dot{R},$ with the upper sign 
corresponding to 
$\alpha<1/2.$ The Allen-Cahn equation has the form 
\begin{equation}
\frac{\gamma}{2K}\dot{R} = \frac{1}{R_*} - \frac{1}{R},
\end{equation}
where 
\[ R_* = \frac{4l_0}{|1-2\alpha|}. \]
Integration of equation (48) yields the following formula
\begin{equation}
\frac{R(t)}{R_*} + \ln|\frac{R(t)}{R_*}-1| = \frac{2K}{R_*^2\gamma} t +
\frac{R(0)}{R_*} + \ln|\frac{R(0)}{R_*}-1|.
\end{equation}

Evolution equation (41) for $C_0$ now has the from
\begin{equation}
\frac{\gamma}{2K} \dot{C}_0 - \frac{1}{R(t)^2} C_0 = \frac{a_1}{2R(t)^2}.
\end{equation}
It has the following solution
\begin{equation}
C_0(t) =-\frac{ a_1}{2} \frac{R_*}{R(t)} 
\frac{ R(0)-R(t)}{R(0)-R_*}
\end{equation}
which obeys the initial condition $C_0(0) =0$.  

Evolution equation for $C_1$ is obtained from the general equation (43). 
For the spherical bubble it has the form 
\begin{equation}
\frac{\gamma}{2K} \dot{C}_1 - \frac{1}{R^2}C_1 = \mp 
2 (\frac{1}{4}a_2 +a_1C_0 +C_0^2) \frac{1}{R^3},
\end{equation}
where as usual the upper sign corresponds to $\alpha <1/2.$ We do not know 
explicit solution of this equation.

There are two cases when the spherical droplets grow: a droplet of the 
ordered phase when 
$\alpha <1/2$, and a droplet of the disordered phase when $\alpha>1/2$.
In both cases formula (46) for the radial velocity of expansion 
$\dot{r}_0$ gives  
\begin{equation}
\frac{\gamma l_0}{2K} \dot{r}_0 =
\frac{|2\alpha-1|}{4} - \frac{l_0}{R} + |a_1(\alpha)| (\frac{l_0}{R})^2
- a_2(\alpha) (\frac{l_0}{R})^3.
\end{equation}
The expansion velocity  is identical  
for all surfaces of constant $\phi$. 

It is clear that there is a minimal R, let us denote by 
$R_{\mbox{min}}(\alpha)$, such that  $\dot{r}_0 >0$. 
We have found numerically that
\begin{equation}
 R_{\mbox{min}}(\alpha) = \frac{ R_*}{ z(\alpha)},
\end{equation}
where the function $z(\alpha)$ is symmetric with respect to $\alpha=1/2$ and 
it has values in the interval [0.866, 1.049]. For example, $z(0) =0.866, \;
z(0.1)=0.984,\; z(0.20)= 1.045,\; z(0.3)=1.040 \; z(0.4)=1.012,\; 
z(0.5)=1.0.$ 
Notice that $R_{\mbox{min}}(\alpha)$ diverges when $\alpha \rightarrow 1/2$.
Thus, in the Ginzburg-Landau model nucleation of expanding droplets is 
possible only if we heat the ordered phase to a temperature above $T_0$, 
or cool the isotropic phase below $T_0$.

For large time $t$, when the droplets are very large, the velocity 
$\dot{r}(t)$ becomes equal to the velocity of the planar interface
\[ \dot{r}_{\infty}(\alpha) = \frac{K}{2\gamma l_0} |1 - 2\alpha|, \]
as expected. 
Notice  that $R_{\mbox{min}}(\alpha)$ and $\dot{r}_{\infty}(\alpha)$
are not independent:
\[ z(\alpha)
\dot{r}_{\infty}(\alpha) R_{\mbox{min}}(\alpha) = \frac{2K}{\gamma }.  \]

Parameter $\alpha$ is related to the temperature:
\[
\alpha = \frac{2\theta}{1-2\theta + \sqrt{1-4\theta}}, 
\]
where 
\[  \theta = \frac{8aC}{9B^2} (T-T_*) \]
is a reduced temperature.  $T_0$ and $T_c$ correspond to 
$\theta=2/9$ and 1/4,  correspondingly. The interval $\alpha\in[0,1/2]$ 
corresponds to $\theta\in [0,2/9]$, and $\alpha\in[1/2,1]$ to $\theta \in
[2/9,1/4]$.  The temperature dependence of $l_0$ can be seen from formula
\begin{equation}
l_0 = (1+\alpha) \frac{\sqrt{2KC}}{3B},
\end{equation}
which follows from the definition (9) after some algebraic manipulations.
$ R_{\mbox{min}}$ is proportional to $l_0/|1-2\alpha|$, 
which can be written in the form
\[
\frac{l_0}{|1-2\alpha|} = \frac{\sqrt{2KC}}{6B} 
\frac{4}{|3\sqrt{1-4\theta}-1|}.   
\]
In the interval $\theta \in [0, 2/9]$, 
which corresponds to $T\in [T_*, T_0]$,  
$l_0/|1-2\alpha| $ monotonically grows from $\sqrt{2KC}/3B$ to infinity. 

Using formula (55) and the symmetry of $z$:  $z(1/2-\delta)=z(1/2+\delta)$,
we obtain the following relation
\begin{equation}
\frac{R_{\mbox{min}}(1/2-\delta)}{R_{\mbox{min}}(1/2+\delta)} =
\frac{3-2\delta}{3+2\delta} <1, 
\end{equation}
where $\delta \in (0,1/2)$.
Thus, the minimal size of the droplets of the isotropic phase which 
appear and grow when $\alpha>1/2$ is significantly larger than the size of 
the droplets of the ordered phase which can appear for $\alpha<1/2$. 

The velocities $\dot{r}_{\infty}$ depend on temperature. In particular, 
comparing them for temperatures below and above $T_0$,
\[  \frac{\dot{r}_{\infty}(1/2 -\delta)}{\dot{r}_{\infty}(1/2 +\delta)}
= \frac{3+2\delta}{3- 2\delta}>1. \]
Hence, the droplets of the isotropic phase  expand more slowly 
than the droplets with $\phi \cong 1$ inside. 

Our main goal in this paper has been to develop the perturbative expansion 
for the curved interfaces. We plan to apply it to interfaces in liquid crystals
in a subsequent work. Nevertheless, just in order to get an idea what 
our formulas predict, we have estimated $l_0$ and $\dot{r}_{\infty}$
for interfaces in nematic liquid crystal MBBA. The model defined by
formulas (1), (2) and Eq.(3) can be related to  de Gennes-Landau theory in a 
single elastic constant approximation ($L_1=K, L_2=0$).    We take data
found in \cite{12,13,14}: $T_*\approx 316K, a\approx 0.021 J/(cm^3\cdot K), 
B\approx 0.07 J/cm^3, C\approx 0.06 J/cm^3 $ (after a change to our 
notation),  and  
$K\approx 6\cdot 10^{-12} N$,  $\gamma \approx 5.2 
\cdot 10^{-2} kg/(m\cdot s)$. We have identified $\gamma$ with the rescaled 
rotational viscosity $\gamma_1 L_1/K_{11}$ at a temperature close to $T_*$. 
Then,  $ T_0 -T_*  \approx 1K,\;\; T_c-T_* \approx 1.2K$. 
The width $l_0$ and the velocity $\dot{r}_{\infty}$ of the planar 
interface are given by  the following formulas:
\[
l_0 \approx 40 (1+\alpha) \cdot 10^{-8}cm, \;\;\; \dot{r}_{\infty}(\alpha)
\approx 1.4 \frac{|1-2\alpha|}{1+\alpha} \frac{cm}{s}.
\]
Notice that even for rather small droplets with the radius of several hundred
\AA ngstr\"{o}m   the ratio $l_0/R$ is rather small.

\section{Remarks}
1. We have shown how one can systematically compute curvature corrections to 
the transverse profile $\phi$ and to the local velocity $\dot{r}_0$ 
of the interface. Due to the presence of functions $C_k$ evolution
equation for the surface $S$ has the relatively simple form (34) and
there are no curvature corrections to it. The formalism for
interfaces is a generalisation of the one constructed for domain walls 
\cite{3,4,5}. The main new ingredients are the $C_0$ function and the 
$(2\alpha-1)/l_0$ term in Allen-Cahn equation (34). By including them we 
have significantly enlarged the range of physical applications of the 
perturbative scheme. This justifies the present publication. 

\noindent
2. The model we have considered is special in the sense that the 
exact planar interface solution $\phi_0$ is known. 
Moreover, the solutions of the equations
$\hat{L}\phi_n = f_n$ are given (almost) explicitly too, because the 
one dimensional integrations in formula (60) below can be easily calculated 
numerically. Here the crucial point is that we know explicitly the two 
linearly independent solutions $\psi_r$ and $\psi_2$ of the homogeneous 
equation $\hat{L}\psi =0$. In other models, the analogs of $\phi_0, \psi_r,
\psi_2$ can be found at least numerically because the pertinent equations
are relatively simple differential equations with the single independent
variable $\xi$ (or $s$ after a rescaling). In our perturbative scheme we 
need to perform explicitly only integrals over $s$, as in integrability
conditions (32) or in formula (60) for $\tilde{\phi}_n$. Such integrals 
can easily be calculated numerically also in the case when only numerical
solutions $\phi_0, \psi_r, \psi_2$ are known.

\noindent
3. In our approach, in order to describe the evolution of curved interface we
introduce the surface $S$, and the functions $C_k$ which can be regarded as
auxilliary fields defined on $S$ and coupled to extrinsic curvatures of it. 
The corresponding evolution equations, that is Allen-Cahn equation (34)for $S$
and equations (41), (43) and analogous ones for $C_k$, have one
independent variable less than the original equation (3). This is significant
simplification from the viewpoint of both computer simulations and analytical
approaches. Therefore we think that our  perturbative scheme is an
interesting tool for investigations of dynamics of interfaces in 
Ginzburg-Landau effective models. 

\noindent
4. The perturbative solution we have presented above is based on the planar
homogeneous interface $\phi_0(s)$. Moreover, the dependence on the transverse
coordinate $s$ is 
uniquely fixed by the perturbative scheme once the initial position of the
surface $S$ and initial data for the functions $C_k$ are fixed. 
This means that in our scheme we obtain a special 
class of interfaces, distinguished by the particular form of the 
dependence on $s$. In other words, the transverse profiles of the 
interfaces provided by the perturbative solution are not arbitrary. 
Intuitively, the interfaces can be regarded as
the planar interface folded to a required shape at the initial instant 
of time, and modified by necessary curvature corrections. Therefore,
it seems appropriate to regard the curved interfaces obtained in our paper 
as the basic ones. More general  interfaces could be obtained by choosing 
more general initial data and solving Eq.(3). For such generic interfaces no
analytic perturbative approach is available. 

\noindent
5. One of advantages of a systematic perturbative approach is that one 
can  make trustworthy estimates of neglected contributions and in consequence
to check whether a given perturbative result is reliable. Formalism with that
level of control can be used in order to make straightforward predictions of
dynamical behaviour of an interface, but perhaps more important application
is to "inverse problems", that is determination of parameters of 
Ginzburg-Landau effective theory. From dynamical behaviour of the interfaces 
one could reliably infer what values have parameters of the model. For 
example,  for a given liquid crystal one could experimentally determine 
coefficients in front of higher powers of the order parameter like 
$\Phi^5, \Phi^6$ or terms of the type 
$\Phi \partial_{\alpha} \Phi\partial_{\alpha} \Phi$ in the formula for 
free energy $F$. For a discussion of the form of $F$ for liquid crystals see,
e.g., \cite{16}.

One could try to generalize our perturbative scheme for calculating the 
curvature corrections to interfaces coupled to a noise. 
In that case the r.h.s. of Eq.(3) would contain an external stochastic force
which in particular would lead to fluctuations of the planar interface. 
In our formalism the dependence of, e.g., the surface tension $\kappa$ on 
curvature radia $R_i$ is explicit and it comes from purely geometric 
quantities like the metric tensor or Jacobian, while numerical coefficients 
in front of powers of $l_0/R_i$ are given by integrals over the $s$ 
coordinate and they are determined essentially by properties of the planar 
interface. Therefore, one may expect that if the stochastic force is present, 
values of these integrals would have to be averaged over the stochastic 
ensemble. This is another interesting direction in which one could continue 
the present work.

\section{Appendix A. The equations $\hat{L}\phi_n = f_n$}

In order to determine $\phi_n$ we have to solve Eq.(26).
Using standard methods \cite{15} it is not difficult to obtain
appropriate solution. 

Let us shift the variable $s$,
\[ s=x+C_0,  \]
in order to remove function $C_0$ from $\phi_0$ present in the
operator $\hat{L}$. Then Eq.(26) acquires the following form
\begin{equation}
\tilde{L} \tilde{\phi}_n(x,\sigma^i,t) = \tilde{f}_n(x,\sigma^i,t),
\end{equation}
where 
\[
\tilde{L} = \frac{1}{2}\frac{\partial^2}{\partial x^2} + (\alpha-\frac{1}{2})
\frac{\partial}{\partial x} -\alpha +2(\alpha+1) \tilde{\phi}_0(x) -
3\tilde{\phi}^2_0(x),
\]
\[
\tilde{f}_n(x,\sigma^i,t) = f_n(x+C_0,\sigma^i,t), \;\;\;
\tilde{\phi}_n(x,\sigma^i,t)= \phi_n(x+C_0,\sigma^i,t),
\]
and
\[
\tilde{\phi}_0(x)= \frac{\exp(x)}{1+\exp(x)}. 
\]

In the first step we find two linearly independent solutions of the
homogeneous equation $\tilde{L}\tilde{\psi}=0.$
The zero mode $\psi_r(x)$ is one solution of this homogeneous equation, 
and the other one has the form
\[
\psi_2(x) = \psi_r(x) h(x),
\]
where $\psi_r=d\tilde{\phi}_0/dx$ and
\begin{eqnarray}
& h(x) = - \frac{1}{2\alpha +1} \exp[-(2\alpha+1)x]- 
\frac{1}{\alpha} 2\exp(-2\alpha x)  &  \\
& + \frac{6}{1-2\alpha} [\exp[(1-2\alpha)x] -1] + 
\frac{1}{1-\alpha}2\exp[2(1-\alpha)x] + \frac{1}{3-2\alpha}
\exp[(3-2\alpha)x]. & \nonumber
\end{eqnarray}
For $\alpha \rightarrow 1/2$ the first term in second line reduces to $6x$. 
Having those solutions one can construct the
relevant Green's function and the solution $\tilde{\phi}_n$,
\begin{eqnarray}
& \tilde{\phi}_n(x, \sigma^i, t) = - 2 \psi_r(x) \int^x_0 dy  
\psi_l(y) h(y) f_n(y, \sigma^i,t)  & \nonumber \\
& + 2 \psi_2(x) \int_{-\infty}^x dy \psi_l(y) 
f_n(y,\sigma^i, t) + C_n(\sigma^i,t)\psi_r(x),  & 
\end{eqnarray}
where 
\[ \psi_l(y) = \exp[(2\alpha-1)y] \psi_r(y). \]
The functions $C_n, n=1,2,...$, are utilised to saturate the integrability 
conditions (32).

The solution of Eq.(26) is given by the formula
\[ \phi_n(s, \sigma^i, t) = \tilde{\phi}_n(s-C_0, \sigma^i, t).  \]

\section{Appendix B. Stability of the interface}

Significance of our theoretical analysis of the interfaces depends on their
stability with respect to small perturbations. It is sufficient to check the
stability of the planar interface because our perturbative solution is based
on it. Mathematically, the stability is related to the sign of eigenvalues
of certain operator, and it is a model dependent property. Considerations
presented below apply to the model defined by formulas (1-3), of course. 
In this case the linearised evolution equation for small amplitude
perturbations $\delta \phi$ of the planar interface $\phi_0(s)$, formula (19),
in the comoving reference frame has the form
\[
\frac{2l_0^2\gamma}{K} \partial_t\delta\phi = \hat{L}\delta\phi 
+ 2l_0^2(\partial^2_{x^1} + \partial^2_{x^2}) \delta \phi,
\]
where $\hat{L}$ has been given below formula (24), and $x^1, x^2$ are the
two Cartesian coordinates in the plane of the interface. Because $\hat{L}$
does not depend on $x^1, x^2,$ we may pass to Fourier modes  
$\tilde{\delta\phi}$ of $\delta\phi$ with respect to these coordinates.
After the substitution
\[ 
\tilde{\delta\phi} = \exp[(\frac{1}{2} - \alpha)s]\Psi
\] 
we obtain the following equation
\[ 
-\frac{2l_0^2\gamma}{K} \partial_t\Psi = \hat{N}\Psi,
\]
where 
\[ 
\hat{N}= -\frac{1}{2} \partial_s^2 + \frac{1}{2}(\alpha +\frac{1}{2})^2
+ 2l_0^2 k^2- 2(\alpha +1) \phi_0 +3\phi_0^2,
\]
with $k^2 = (k_1)^2 + (k_2)^2, \; k_1, k_2$ being the wave numbers Fourier
conjugate with $x^1, x^2$. 

Notice that 
\[ 
\Psi_0 = \exp[(\alpha -\frac{1}{2})s] \psi_r,
\]
where $\psi_r$ is the zero-mode given by formula (29), is eigenfunction of
the Hermitean operator $\hat{N}$. The corresponding eigenvalue is equal 
to zero. Because $\Psi_0$ 
does not vanish at any finite $s$, it represents "the ground state" of a
fictitious system with $\hat{N}$ as "the  Hamiltonian". Hence, all other 
eigenvalues of $\hat{N}$ are strictly positive. Moreover, looking at 
"the potential" in "the Hamiltonian" $\hat{N}$ one can see that the zero
eigenvalue and the next one are separated by a finite gap. The eigenmode 
$\Psi_0$ corresponds to a parallel shift of the interface as a whole in 
the $x^3$ direction. All other eigenmodes decay exponentially with 
a characteristic time equal to
$t_c= 2l_0^2\gamma/(K\lambda)$, where $\lambda$ denotes the corresponding
eigenvalue. Thus, the planar interface in our model is stable. Consequently,
also the curved interfaces are stable with respect to small perturbations
provided that their curvature radia are large enough for validity of our
perturbative expansion. 

Let us point out that from point of view of applications in condensed
matter physics even unstable interfaces can be interesting if unstable
modes grow in time so slowly that the interface manages to travel across 
the sample before these modes become visible. The interface has
finite normal velocity $\vec{p}\dot{\vec{x}}_0$, formula (46), and in any 
real experiment the sample occupies a finite volume.

\end{document}